\documentclass[11pt,twoside]{article}
\usepackage{asp2004}
\usepackage{psfig}
\usepackage{epsf}
\usepackage{graphics}
\usepackage{lscape}
\def\kpc{\,{\rm kpc}}

\def\Myr{\,{\rm Myr}}\def\Gyr{\,{\rm Gyr}}
\def\msun{\,{\rm M}_\odot}
\def\fracj#1#2{{\textstyle{#1\over#2}}}

\def\edcomment#1{\iffalse\marginpar{\raggedright\sl#1\/}\else\relax\fi}
\reversemarginpar

\begin{document}
\title{The Cosmological Context of Extraplanar Gas}
\author{James Binney}
\affil{Theoretical Physics, Keble Road, Oxford, OX1 3NP, England}

\begin{abstract}
I review evidence that galaxies form from gas that falls into potential
wells cold, rather than from virialized gas, and that formation stops once
an atmosphere of trapped virialized gas has accumulated. Disk galaxies do not
have such atmospheres, so their formation is ongoing.  During galaxy
formation feedback is an efficient process, and the nuclear regions of disk
galaxies blow winds. The cold infalling gas that drives continued star
formation has a significant component of angular momentum perpendicular to
that of the disk.  Extraplanar gas has to be understood in the context set
by nuclear outflows and cold skew-rotating cosmic infall.
\end{abstract}
\thispagestyle{plain}

\section{Introduction}

For a quarter of a century work on galaxy formation has been dominated by
the belief that when a density perturbation goes non-linear and collapses,
its baryons are shock heated to the virial temperature $T_{\rm vir}$, so the
raw material for galaxy formation is hot gas, and galaxies form by the
cooling of virial temperature gas \citep{ReesOstriker,WhiteRees}. Several
recent developments indicate that this picture is fundamentally flawed
\citep{Binney04a}. In the process of summarizing these developments I hope
to persuade you that galaxies form from gas that is much colder than $T_{\rm
vir}$, and that galaxy formation ceases once all its gas becomes hot.
Feedback is an efficient process, and Active Galactic Nuclei (AGN) play a
crucial role by setting the upper limit to the luminosities of galaxies. In
disk galaxies a wind from the nuclear regions coexists with infall of cold
gas to the disk, and the dynamics of extraplanar must be affected by
interactions with both outflowing and infalling material.

\section{Lessons from `cooling flows'}

The deep gravitational potential wells of massive elliptical galaxies and
clusters of galaxies have long been known to contain major quantities of
virial-temperature gas that can be studied through its thermal X-ray
emission. Until recently it was conventional to suppose that this gas is
passively cooling \citep[e.g.][]{fabian95}. To reconcile this belief with
the measured radial surface-brightness profiles of these systems, it was
necessary to conjecture that the medium was everywhere multiphase, and that
throughout the sphere in which the cooling time $t_{\rm cool}=3kT/2\dot E$
equals the Hubble time, very cold gas was `dropping out'. X-ray spectra from
the new satellites show that in clusters of galaxies there is much less gas
at $T\la \fracj13T_{\rm vir}$ than the cooling-flow model predicts
\citep{Bohringeretal,Petersonetal}. In fact, there is no spectroscopic evidence for
temperature variations on very small scales within the X-ray emitting gas --
the gas appears not to be multiphase after all.  There \textit{are}
temperature fluctuations on scales large enough to be resolved by
\textit{Chandra}. Downward fluctuations are associated with the filaments of
cold gas that have long been a puzzle. The strong H$\alpha$ emission from
these objects is probably powered by embedded star formation, and  smooth
temperature gradients appear to connect a filaments to the embedding thermal
plasma \citep{Fabianetal}. Elsewhere sharp steps in $T$ are observed that
extend large distances through the source \citep{Markevichetal}. These `cold
fronts' are presumably contact discontinuities of the type expected in a
highly turbulent fluid.

An AGN invariably sits at the centre of a `cooling flow', and there is now
clear evidence that jets thrown out by this object are doing significant
work on the thermal plasma \citep[e.g.][]{NulsenDMJFW,Heinz}. The jets create regions of low or
negligible thermal X-ray emission, and weak shock waves must be spreading
out from these `cavities'. Transsonic turbulence must be generated as the
underdense cavities rise and break up 
\citep{Chirazov,Quilis,ChurazovSFB,Kaiser,Ommaetal,Omma}.  These
phenomena confirm the correctness of the general picture I developed with
Gavin Tabor a decade ago \citep{TaborB,BinneyT} in which an unsteady
equilibrium is set up between radiative cooling and heating through the
temperature dependence of the accretion rate onto the central black hole. We
argued that the dominant channel for the heating is the decay of the
turbulence that is driven by jets emanating from the black hole. Although
some controversy still surrounds this point \citep{Birzanetal}, the data appear
to be compatible with our prediction \citep{Binney03,NipotiB04}.

What needs stressing is the implication of these findings for the
White--Rees picture of galaxy formation: where we actually see trapped
virial-temperature gas with a short cooling time, it isn't cooling to form
stars, but seems to be thermostated at $T\simeq T_{\rm vir}$. What reason
have we to suppose that gas at $T_{\rm vir}$ behaved differently in the
past? 

\section{Lessons from quasars and stellar populations}

It is now clear that quasars and other AGN are powered by the formation of
the black holes (BHs) that are found at the centres of all sufficiently
nearby, luminous
galaxies. Studies of the demographics of these BHs and the luminosity
functions of AGN of various types in different redshift ranges give strong
indications of the role that BHs play in galaxy formation. Two findings are
crucial:

\begin{itemize}
\item There is a tight correlation between the mass of the galactic-centre
BH and the velocity dispersion of the host spheroid, which is itself
correlated with the spheroid's mass \citep{Tremaineetal}.

\item The energy radiated by luminous quasars is comparable to the energy that
can plausibly be extracted during BH formation.  Moreover, most of this
energy must have been radiated at near-Eddington luminosities
\citep{YuTremaine}. Consequently, the characteristic timescale on which BHs
grow is the Salpeter time $t_{\rm Salt}\simeq25\Myr$, and a BH can grow from
a seed mass $\la1000\msun$ to the largest observed BH masses $\sim10^9\msun$
in a time $\la1\Gyr$ that is small compared to the Hubble time.

\end{itemize}

Comparison of the redshift evolution of the quasar density and the density
of star formation then leads to the conclusion that

\begin{itemize}

\item The rate of black-hole formation is consistent with being proportional to
the rate of overall star formation \citep[e.g.][]{Haimanetal}.

\end{itemize}

Finally, studies of early-type galaxies at low redshift \citep{Sauron}
show that 

\begin{itemize}
\item the central regions of massive spheroids are old and have enhanced
abundances of the $\alpha$ elements, implying that they formed all their stars
at significant redshift and on a timescale that is short compared with the
timescale $\ga1\Gyr$ for enrichment by type Ia supernovae.

\end{itemize}

From these facts is clear that BHs and spheroids formed together during
orgies of star-formation and topsy BH growth. This era of growth may have
been broken into several episodes, but the elapse of time from the onset of
serious star formation through its completion cannot have significantly
exceeded a Gyr. 

Since stars are observed to form from very cold ($T\la30\,$K) gas, it
follows that BHs fed from such cold gas also.  Presumably they fed quickly
at this stage because they fed off very dense cold gas. Now they are growing
very much more slowly because they can eat only hot rarefied gas. Then the
energy released as they grew was radiated with great efficiency by the dense
surrounding gas. Now that they are surrounded by plasma too rarefied to
radiate efficiently, jets carry away most of their energy output
\citep{Owen,Matteo}. In fact, it is not clear that BHs are growing at all at the
present epoch, since the energy carried away by the jets may be energy
stored in BH spin since those early days of topsy growth.

\section{The critical mass $M_*$}

We have seen that there are observational indications that spheroids formed
from cold gas rather than gas at $T_{\rm vir}$ as \cite{WhiteRees} assumed.
Simulations of galaxy formation suggest a substantial fraction of the
baryons that fall onto a protogalaxy should, in fact, arrive cold rather than
at $T_{\rm vir}$. \cite{Katzetal} used N-body/SPH simulation to determine as
a function of redshift the rate of infall of gas onto a typical protogalaxy,
and for each parcel of gas determined the highest temperature $T_{\rm max}$
achieved. They found that the rate of infall declined with $z$, slowly from
$z=3$ to $z=2$, and then quite rapidly. At all redshifts $T_{\rm max}$ ranged
from $10^4\,$K to $10^7\,$K. At $z=3$ the distribution of $T_{\rm max}$ was
strongly bimodal with about equal quantities arriving either side of
$2\times10^5\,$K. At the present epoch the distribution was fairly flat,
with a continuous transition at intermediate $z$.

\cite{BirnboimD} studied the infall of gas onto a galaxy-sized overdensity
semi-analytically. They showed that post-shock cooling is so efficient that
the accretion shock does not break away from the centre of the system in
systems that contain $\la3\times10^{10}\msun$ of baryons. This mass
coincides remarkably accurately with the characteristic baryonic mass $M_*$ that
emerges from the SDSS data for relatively nearby galaxies
\citep{Kauffmannetal}: below $M_*$ surface brightness increases with $M$,
the galaxies are relatively young and not very centrally concentrated, while
above this mass surface brightness is independent of $M$, and galaxies are
old and centrally concentrated.

It has been recognized for a very long time \citep{Larson,DekelS} that
heating by core-collapse supernovae picks out a mass scale that is similar
to $M_*$: for a given IMF, supernovae inject a well defined energy per unit
mass. The temperature $T_{\rm SN}$ to which the ISM can be heated by this
energy depends on the efficiency of radiative cooling and on how uniformly
the energy is distributed. A reasonable estimate is obtained by neglecting
radiation losses (which leads to $T_{\rm SN}$ being overestimated) and
assuming that the energy is shared by all the surviving gas (which causes
$T_{\rm SN}$ to be underestimated).  With these assumptions, $T_{\rm
SN}\sim1\,$keV. $T_{\rm SN}=T_{\rm vir}$ for $M\sim M_*$.

Hence it is unclear whether $M_*$ has been imprinted on the observational by
the physics of gravitational heating, or by SN heating. However, the
following line of argument suggests that SN heating is the more important
factor. 

In all systems at early times there will be an abundance of cold gas and
star formation. SNe will quickly heat pockets of gas to $T_{\rm SN}$. If
$M<M_*$ this gas will escape into the intergalactic medium (IGM), carrying
off the bulk of the SN energy and much of the newly-synthesized elements.
In clusters the significant metallicity of the X-ray emitting gas (which
contains $\sim4/5$ of the baryons and $\sim1/2$ of the metals) is direct
evidence that this happened. In systems with $M<M_*$ gravitational heating is
ineffective, so alongside the outflow of SN-heated gas, there will be an
inflow of cold metal-poor gas that continues to feed star formation. 

Once $M>M_*$, the SN-heated gas cannot escape, although it can be pushed out
of the region of most active star formation. There it will be joined by the
now significant fraction of gravitationally heated gas. Over time a `cooling
flow' will develop. But, as we now know, the central BH will ensure that no
stars form from this virialized gas. Cold gas continues to fall into the
system, but, as in the Katz et al.\ simulations, at an ever-diminishing
rate. Cold gas must enter as blobs and streamers; presumably blobs are often
tidally disrupted into streamers. The orbits of infalling blobs will be
determined by their angular momentum, which will on average increase over
time, as material arrives that started out more and more remote from the
centre of the system. Hence the fraction of the infalling blobs that
penetrate the dense core of the cooling flow decreases over time. 

The coexistence of hot and cold gas in pressure equilibrium is unstable: if
the pressure is high, conduction of heat into the efficiently radiating cold
gas will cause a steady transfer of gas from the hot to the cold phase,
whereas if the pressure is low and radiation less efficient, the conductive
heat flux will cause the cold gas to evaporate and join the hot phase. At
each pressure there is a critical filament length $l_{\rm crit}$ above which
filaments condense hot gas and below which they are evaporated by
conduction.  \cite{NipotiB} showed that in the very centres of cooling-flow
clusters, where cold filaments are observed, $l_{\rm crit}\sim10\kpc$, while
elsewhere $l_{\rm crit}$ is so large that it is not surprising that there
are no surviving filaments. Hence this line of argument not only allows us
to explain why we only see H$\alpha$ filaments very close to the centres of
cooling-flow clusters, but also explains why star formation is confined to
these regions also: the atmosphere of gas at $T_{\rm vir}$ cuts off the
supply of cold gas by transferring to infalling protogalactic material
energy generated by the nuclear BH. The fact that filaments contain normal
dust \citep{SparksEtal89}, which condensed virial-temperature gas would not,
supports the role of cold infall in the formation of filaments.

\section{The mass function of DM halos and the luminosity function of
galaxies} 

Although in conventional CDM cosmology all galaxies form by the accumulation
of baryons in DM halos, the luminosity function of galaxies is profoundly
different from the mass function of DM halos. The latter is very nearly a
power law $dN/dM\propto M^{-2.17}$ in the mass range of interest, while the
luminosity function of galaxies is a much flatter power law for $L<L_*$ and
then cuts off with almost exponential steepness. The characteristic
luminosity $L_*$ coincides rather exactly with the luminosity of the Milky
way, which has a baryonic mass $\sim M_*$. From the earliest days of the CDM
theory it has been clear that the steep slope of the DM mass function must
be reconciled with the flat slope of the luminosity function by the
efficient ejection of gas from shallow potential wells. SN-driven winds are
prime candidates for this job, but gas is probably removed from the
shallowest wells at very early times by photoelectric heating
\citep{Efstathiou} and thermal conduction \citep{Dekel}.  Ab initio
simulations of galaxy formation in the CDM picture have struggled to produce
efficient mass ejection, but this is generally ascribed to inadequate
spatial resolution, and the semi-analytic galaxy-formation codes that are
widely used to connect observations to N-body simulations of invisible DM
assume that `feedback' from star formation to the ISM is efficient.

 \cite{Bensonetal} show that in the convergence $\Lambda$CDM cosmology
feedback of the required efficiency is problematic in that it leads to the
formation of excessive numbers of very luminous and anomalously young
galaxies: gas thrown out of small halos later finds its way into deep
potential wells, where it cools to form stars. This problem is eliminated
once we recognize that these deep potential wells are precisely those
possessed of AGN-thermostated `cooling flows'. Hence the late-infalling
baryons that plague the Benson et al.\ models do not lead to the formation
of ultraluminous galaxies, but simply extend the X-ray emitting atmospheres
of groups and clusters.

\section{Formation of disk galaxies}

This meeting is about gas found away from the planes of disk galaxies. So
how do disk galaxies fit into this picture?

The age distribution of stars near the Sun indicates that stars have formed
at an approximately steady rate for $\ga10\Gyr$ \citep{BinneyDB,Nordstrom}.
This phenomenon suggests that the solar neighbourhood has been accreting gas
steadily over this period -- if it were simply using an initial stock at a
steady rate, not only would there be many more metal-poor G-dwarf stars than
are observed, but it would be hard to understand why the strongly
Jeans-unstable gas-rich early disk did not form stars much more rapidly than
the relatively stable gas-poor current disk does. 

Given the evidence presented above, it seems reasonable to assume that most
of the infalling gas arrived cold rather than gravitationally heated to
$T_{\rm vir}$. In support of this assumption we have the result of
\citep{Frenketal} that the contrary assumption of White \& Rees leads to
predicted soft X-ray luminosities of edge-on disk galaxies that are factors
of several too large.

We do not understand how gas joins the disk, but we have good reason to
believe that in general it arrives with an angular-momentum vector that is
inclined to the disk's symmetry axis, specifically:

\begin{itemize}

\item One explanation of the ubiquitous
warps of HI disks (and the rarer warps in stellar disks) is that the torque
generated by the deviation from planarity is associated with the exchange
of angular momentum between the inner disk and the periphery, the angular
momentum of which is dominated by contributions from recently accreted
material \citep{OstrikerB,JiangB}. Although we still lack a clinching
argument for the correctness of this theory, all competing theories of warps
have been ruled out.

\item The angular momentum in the Magellanic Stream is comparable to that in
the Galactic disk, and the two vectors are nearly perpendicular. The
Sagittarius Dwarf has an intermediate stellar population so it must have
started out in possession of interstellar gas. This has by now been stripped
and will have contributed angular momentum in the direction that is
perpendicular to both the disk and the orbit of the Magellanic Stream.

\item Polar ring galaxies sometimes have much more angular momentum in the
ring than in the perpendicular disk \citep{Iodice}.

\item
The famous galaxy NGC 4550, which has two coextensive stellar disks that
rotate in opposite senses \citep{Rubinetal}, must have formed through the
accretion by a disk galaxy of gas that had angular momentum in the opposite
direction to the galaxy's existing angular momentum.

\item
The standard theory of overdensities in a Gaussian random field predicts
that the angular momentum vector of infalling matter changes direction by of
order a radian in a Hubble time \citep{QuinnB}.

\end{itemize}

Infall of material with inclined angular momentum will clearly generate
extra-planar gas.

There is now compelling evidence that the inner part of the Milky Way's
bulge is blowing a wind \citep{BlandH}. This is as expected for a spheroidal
system with $M<M_*$. Somewhere there must be a transition between a wind
from the nucleus and infall to the disk. Moreover there is likely to be mass
and energy exchange between the two flows. As evidence of this exchange, one
may cite the metallicities of high-velocity clouds \citep{Wakker}. The
available measurements indicate values of $Z\sim0.2Z_\odot$ that are much
lower than the expected metallicity ($Z>Z_\odot$) of the nuclear wind, but
higher than the expected metallicity ($Z\sim0.01Z_\odot$?) of primordial gas.

\section{Conclusions}

We have seen evidence (i) that feedback is efficient, (ii) that the nuclear
regions of spirals are blowing winds, (iii) that galaxies form from cold
infall, (iv) that the disks of spirals are still forming, (v) by acquiring
gas that has angular momentum that is not parallel to the disk's angular
momentum. When we put these pieces together it is clear that we expect there
to be extraplanar gas, and that we cannot hope to understand the dynamics of
this gas without taking into account the way in which it interacts both with
the nuclear wind and with cosmic infall. Conversely, studies of extraplanar
gas provide leverage on the central problem of understanding how the
Universe came to be structured as it is.

\end{document}